\documentclass[twocolumn,aps,showpacs,superscriptaddress,prl]{revtex4}
\usepackage{graphicx}
\usepackage{dcolumn}
\usepackage{bm}

\begin{document}

\title{
Tailoring the Magnetic Anisotropy in CoRh Nanoalloys: Experiment and Theory
      }
\author{M.~Mu\~noz-Navia}
\altaffiliation{
Present address: Max-Planck-Institut f\"ur Mikrostrukturphysik,
Halle, D-06120, Germany.
              }
\author{J.~Dorantes-D\'avila}
\affiliation{
Instituto de F\'{\i}sica, Universidad Aut\'onoma de San Luis
Potos\'{\i}, 78000 San Luis Potos\'{\i}, Mexico
           }
\author{D.~Zitoun}
\author{C.~Amiens}
\affiliation{
Laboratoire de Chimie de Coordination, CNRS, 
31077 Toulouse, France
           }
\author{N.~Jaouen}
\author{A.~Rogalev}
\affiliation{
European Synchrotron Radiation Facility, 6 rue Jules Horowitz,
38043 Grenoble, France
           }
\author{M.~Respaud}
\affiliation{
Universit\'e de Toulouse, LPCNO, INSA, 135 avenue de Rangueil,
F-31077 Toulouse, France
           }
\author{G.~M.~Pastor}
\affiliation{
Institut f\"ur Theoretische Physik, Universit\"at Kassel,
34132 Kassel, Germany
           } 
\date{\today}
\begin{abstract}

The magnetic moments and magnetic anisotropy energy (MAE) of 
CoRh alloy nanoparticles are determined experimentally 
and theoretically. Non-trivial correlations between chemical
order, magnetic order and MAE
are revealed. A remarkable non-monotonous dependence of 
the MAE as a function of composition and chemical order is observed that 
opens novel possibilities of tuning the magnetic properties
of nanoalloys. The observations are successfully compared and analyzed
with our electronic calculations. In this way we clearly demonstrate 
that the induced $4d$ moments and the $3d$-$4d$ interfaces are 
the key parameters controlling the magneto-anisotropic behavior.

\end{abstract}
%

\pacs{
75.75.+a, 
36.40.Cg, 
75.30.Gw, 
75.50.Cc  
}
 
\keywords{
Magnetic alloy nanoparticles, magnetic anisotropy, 
spin and orbital moments, CoRh nanoparticles 
}
\maketitle

Recent progress in the growth of multicomponent nanoparticles (NPs) 
with controlled size and composition has opened new routes 
in fundamental and applied materials research 
\cite{sun,prl-zitoun,bansmann}. {\em Nanoalloys} 
are thus emerging as an important multidisciplinary field with
increasing impact throughout nanoscience and technology
\cite{nanoalloys}.  
Many-body phenomena like magnetism are particularly interesting in this 
context, since they are very sensitive to the local and 
chemical environment of the atoms. 
Pure Fe$_N$, Co$_N$ and Ni$_N$ clusters are known to be magnetic 
with stronger spin and orbital moments than the corresponding 
solids~\cite{billasetal}. However, their magnetic anisotropy energy (MAE) 
per atom ---despite being orders of magnitude larger than in 
solids~\cite{prl-mae-95}--- remains relatively 
small due to the relatively weak spin-orbit (SO) coupling in the $3d$ series. 
Stronger SO interactions are certainly available in the heavier 
$4d$ and $5d$ TMs but pure NPs of these elements are non-magnetic, 
except eventually for extremely small sizes 
(e.g., Rh$_N$ for $N\lesssim 30$ atoms) \cite{cox}. 
In view of these contrasting behaviors one concludes that
$3d$-$4d$ and $3d$-$5d$ nanoalloys should show unique magnetic 
properties as a function of size, composition and chemical order,
which remain for the most part to be discovered.
In particular it should be possible to control and optimize the 
spin and orbital moments, their order and magnetic anisotropy 
by varying the $3d/4d$ or $3d/5d$ content as well as the distribution
of the components within the clusters. This is very important, 
specially for applications, since the MAE determines the lowest-energy
magnetization direction as well as its stability with 
respect to temperature fluctuations and reversal processes. It is 
therefore quite remarkable, the intense research activity 
in nanomagnetism notwithstanding, that very little is still known 
about the microscopic origin of the MAE in nanoalloys and about the 
possibilities of systematic material optimization that it 
offers \cite{bansmann,nanoalloys}. Understanding the trends governing
the properties of binary NPs is therefore an important 
step towards quantum-based magnetic nanoalloy design. 
It is the goal of this letter to demonstrate how the magnetic
moments and anisotropy of $3d$-$4d$ nanoalloys can be tailored 
and to reveal the microscopic mechanisms that underlie their 
novel behavior.  

The NPs have been synthesized in solution by decomposition of organometallic 
precursors in the presence of a stabilizing polyvinylpyrrolidone polymer 
as described in Ref.~\cite{prl-zitoun}. This approach allows us to easily 
vary the Co concentration 
$x$ by controlling at the same time the aggregation and the particle 
diameter $\phi$, typically in the range 
$1.6$~nm~$ \le \phi \le 2.5$~nm. For the present study
samples with $x = 0.76$, $0.49$ and $0.25$ were
synthesized. The structural characterization has been done 
using high-resolution transmission electronic microscopy (TEM) 
and wide angle X-ray scattering (WAXS) techniques \cite{prl-zitoun,fromen}. 
TEM results show a regular dispersion of the clusters inside the matrix and 
a narrow log-normal-like size distributions peaked at an average 
$\phi \simeq 2$ nm and having a width below 15\%. 
The WAXS pattern of Co$_{0.49}$Rh$_{0.51}$ and Co$_{0.76}$Rh$_{0.24}$ NPs 
is well fitted with a compact  
structure with nearest neighbor (NN) distance 
$d_{\rm NN}\simeq  0.269$ nm for Co$_{0.49}$Rh$_{0.51}$ and 
$d_{\rm NN}\simeq 0.263$ nm for Co$_{0.76}$Rh$_{0.24}$ \cite{fromen}. 
The Co$_{0.25}$Rh$_{0.75}$ NPs show a bulk-like fcc structure with 
$d_{\rm NN}\simeq  0.269$ nm. 
The fact that $d_{\rm NN}$ is very close to the NN distance
of bulk Rh [$d_{\rm NN}({\rm Rh}) = 0.273$ nm]
and the contrast observed in high-resolution TEM images 
of larger isolated particles indicate that the NPs
are all bimetallic with close-packed structures having
most probably a Rh-rich inner core and a Co-rich outer shell.
The magnetization of the NPs has been measured by SQUID magnetometry.
At low temperatures ($T=2$~K) all systems are found to be magnetic with 
a ferromagnetic hysteresis behavior. As a general tendency, we observe that
increasing the Rh concentration increases the coercive field
as well as the irreversible field even beyond $5$~T. 
The magnetization per CoRh unit $\overline\mu_{\rm CoRh}$ at $5$~T 
first increases and then decreases with increasing Rh concentration: 
$\overline\mu_{\rm CoRh} = 1.83$, $2.2$, $1.9$ and $0.8 \mu_B$ for 
$x = 1$, $0.76$, $0.49$ and $0.25$ in Co$_x$Rh$_{1-x}$ NPs ($\phi = 2$~nm).
The differential high-field susceptibility is also enhanced and
in particular for $x=0.49$ the magnetization curves 
are difficult saturate. We have therefore performed measurements 
in pulsed high magnetic fields up to $30T$ and obtained a proper 
saturation for $x = 0.49$ with 
$\overline\mu_{\rm CoRh} = (2.38 \pm 0.05) \mu_{\rm B}$~\cite{prl-zitoun}.
   
While the basic idea behind tailoring the NP magnetic properties
by alloying $3d$ and $4d$ elements is relatively simple, 
its practical realization involves a number of serious challenges. 
For instance, different growth conditions can lead to segregated clusters 
with either a $4d$ core and a $3d$ outer shell or vice versa. 
Post-synthesis manipulations can induce different degrees of intermixing 
at the $3d$/$4d$ interfaces including surface diffusion, ordered 
or disordered alloys, etc.. Controlling the distribution of the 
elements within alloy clusters is therefore an important experimental issue. 
In order to quantify the role of chemical 
order on the magnetic properties of CoRh nanoalloys we have performed
a comprehensive set of electronic calculations for clusters 
having $N\le 500$ atoms, Co concentrations $x\simeq 0.25$, $0.5$ and 
$0.75$, and different distributions of the Co and Rh atoms. 
The theoretical framework is given by a realistic $spd$-band 
tight-binding Hamiltonian that takes into account the effects of 
redistributions of the spin- and orbital-polarized density, as well as
the SO interactions $H_{SO}$ at the origin 
of magnetic anisotropy \cite{prl-mae-95,ginette}. 
Since the SO interactions are very sensitive 
to the details of the electronic spectrum, we perform accurate 
self-consistent calculations 
for each orientation $\delta$ of the spin 
magnetization $\vec S$. The MAE $\Delta E_{\delta\gamma} = E_\delta - E_\gamma$
is derived in a nonperturbative way as the difference between the electronic 
energies $E_\delta$.  
\begin{figure}
\includegraphics[scale=0.32]{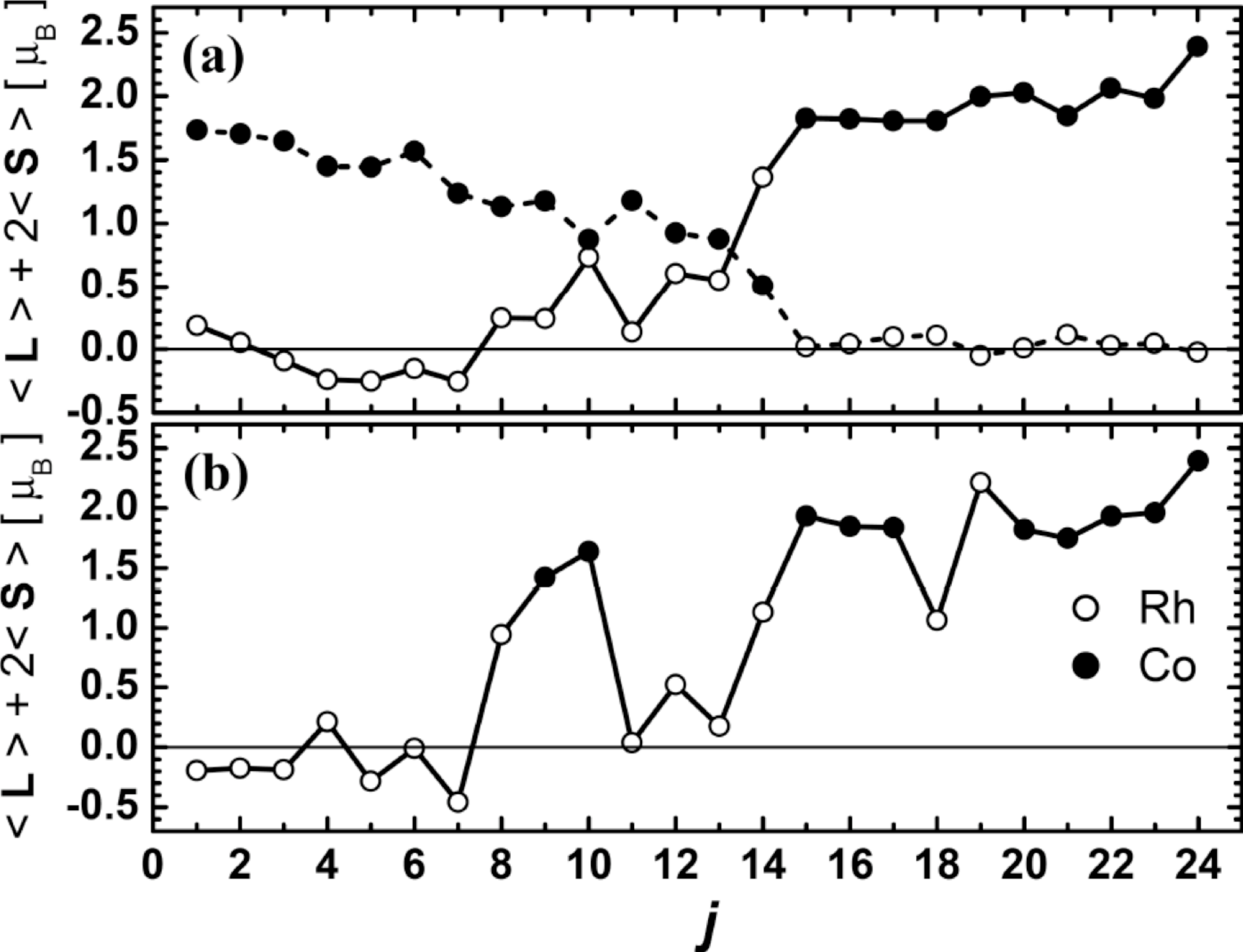}
\caption{\label{fig:moms}
Local magnetic moments $\mu(j) = 2 \langle S(j)\rangle + \langle L(j) \rangle$ 
along the easy axis of 2~nm ${\rm Co}_{0.5}{\rm Rh}_{0.5}$ clusters 
having an fcc-like octahedral structure ($N=489$ atoms).
The atomic shells $j$ are ordered by increasing distance to the cluster center.
(a) Fully segregated Co core and Rh outer shell (dashed) and fully segregated Rh core and
Co outer shell (solid). (b) Rh core and Co outer shell with 
intermixing at the interface. Dots (open circles) refer
to Co (Rh) atoms. 
        }
\end{figure}

In Fig.~\ref{fig:moms} theoretical results are given for the magnetization 
profile of 2~nm octahedral Co$_{0.5}$Rh$_{0.5}$ NPs with three different 
chemical orders ($N=489$ atoms). For simplicity, a compact fcc-like 
structure with bulk Co-Co and Rh-Rh interatomic distances and 
epitaxial interface growth is assumed. 
These results are representative of a much larger number of studied 
structures, shapes, sizes and compositions \cite{theo2b}. In the case 
of a fully segregated Co core with a Rh outer shell 
[dotted curve in Fig.~\ref{fig:moms} (a)] 
the Co magnetic moments at the innermost atoms are similar 
to the Co bulk moment and decrease as one goes 
from the center to the CoRh interface. Here only very small magnetic moments 
are induced at the Rh due to the proximity of the spin-polarized Co. 
As we move away from the CoRh interface towards the surface 
(larger $j$ in Fig.~\ref{fig:moms}) the Rh polarization shows some oscillations. 
Notice that the reduction of the Co moments at the interface is 
not compensated by the tiny induced Rh moments.
Consequently, the calculated average moment per 
CoRh unit $\overline\mu_{\rm CoRh} = 1.30\mu_B$, with spin and 
orbital contributions $\mu_S = 1.21 \mu_B$ and $\mu_L = 0.09 \mu_B$, 
remains fairly small, actually smaller than in pure Co clusters or surfaces. 

The situation changes qualitatively if one considers 
a Rh core with a Co outer shell 
[dashed curve in Fig.~\ref{fig:moms} (a)]. In this case the Co 
moments are largest at the cluster surface ($19\le j \le 24$) and 
decrease only slightly at the CoRh interface. The induced Rh moments
are quite important in particular at the interface 
[$\mu(j) = (0.5$--$1.3)\mu_B$ 
for $10\le j \le 14$]. They yield, despite reductions and changes of sign
at the innermost shells ($1\le j \le 9$), a significant positive
contribution to the CoRh-unit moment $\overline\mu_{\rm CoRh} = 2.08\mu_B$,
which is now larger than in bulk alloys of similar concentrations.
The spin and orbital moments amount here to
$\mu_S = 1.75 \mu_B$ and $\mu_L = 0.33 \mu_B$. 
These contrasting behaviors demonstrate the crucial role of chemical 
order on the magnetic properties of nanoalloys.
In fact, only one of these arrangements, namely a Rh-core with an outer 
Co shell, yields results that are qualitatively consistent 
with the experimental ones, though somewhat smaller 
[$\overline\mu_{\rm CoRh}^{\rm expt} = (2.38 \pm 0.05)\mu_B$ 
for 2~nm Co$_{0.5}$Rh$_{0.5}$].

The trends as a function of chemical order
can be understood by contrasting the different local environments 
of the Co and Rh atoms. In the first case (Co core and Rh shell) 
all Co atoms have bulklike coordination and the Co atoms at the
interface have less Co than Rh NNs. This increases 
the effective local $d$-band width at Co atoms and reduces the local 
Co moments. Moreover, the interface Rh atoms have few Co NN with 
weakened moments, so that the polarization induced by the proximity 
with Co is quite small. Finally, the curvature at the surface 
of 2~nm particles ($N\simeq 500$) is not large enough to sustain
the formation of local Rh moments on the basis of a reduction of the 
local coordination number alone (Rh$_N$ is magnetic only for 
$N\le 30$--$50$ \cite{cox}). In contrast, in the second case (Rh core and
Co shell) there are several factors that {\em enhance} 
magnetism: i) The reduction of coordination number at the 
surface Co atoms increases the local moments, in particular the
orbital contribution. ii) The interface Co atoms, being outside, 
have more Co than Rh NNs, so that the $d$-band broadening  
due to hybridizations with Rh atom is weaker.
iii) The Rh atoms at the interface have here a majority of 
strongly magnetic Co atoms as NNs, that induce important Rh  
moments over several interatomic distances. These trends 
are common to all studied compact clusters with different 
surface shapes and sizes ($N \ge 100$) \cite{foot:rh-regimes}. 
One concludes that in the technologically most interesting NPs, 
which have a few hundreds of atoms, the $4d$ magnetism and the 
associated MAEs can only survive close to the $3d$-$4d$ interfaces. 
The shape and structure of the interfaces and the possible 
interactions with the cluster surface become therefore crucial 
for the magnetic behavior of nanoalloys. 

Real NPs are not expected to have perfect interfaces with completely 
segregated species. We have therefore investigated more complex arrangements 
of the Co and Rh atoms by considering various intermixed
configurations at the interface, 
surface segregation, as well as completely random-alloy arrangements.
A representative example is shown in Fig.~\ref{fig:moms} (b). 
This corresponds to a Rh core and a Co outer shell with
intermixing of the Co and Rh atoms on four layers ($\pm 2$) 
around the interface. One observes an important enhancement 
of both spin and orbital Rh moments, particularly for those Rh 
atoms having a Co-rich local environment, while the 
Co moments are not much affected. 
Basically small reductions are seen at Co atoms in Rh-rich 
environments. The overall result is a significant increase 
of the average magnetic moment of the NPs that improves the 
agreement with experiment
($\overline\mu_{\rm CoRh} = 2.24 \mu_B$ with $\mu_S = 1.90 \mu_B$ 
and $\mu_L = 0.34 \mu_B$ for $N=489$ atoms). 
A similar behavior is also found in the case of random 
distributions of Co and Rh atoms.
The enhancement of $4d$ moments due to intermixing is explained 
by the large spin and orbital polarizability
of Rh atoms with increasing number of Co NNs and the 
robustness of the nearly saturated Co spin moments. A particularly strong
increase of the local moments is observed for Rh atoms 
at the cluster surface, since the contributions 
from Co NNs and reduced coordination number add up. In fact, 
from the point of view of the magnetic moments, Rh atoms 
surrounded mainly by Co NNs behave almost like Co atoms.
\begin{figure}
\includegraphics[scale=0.17]{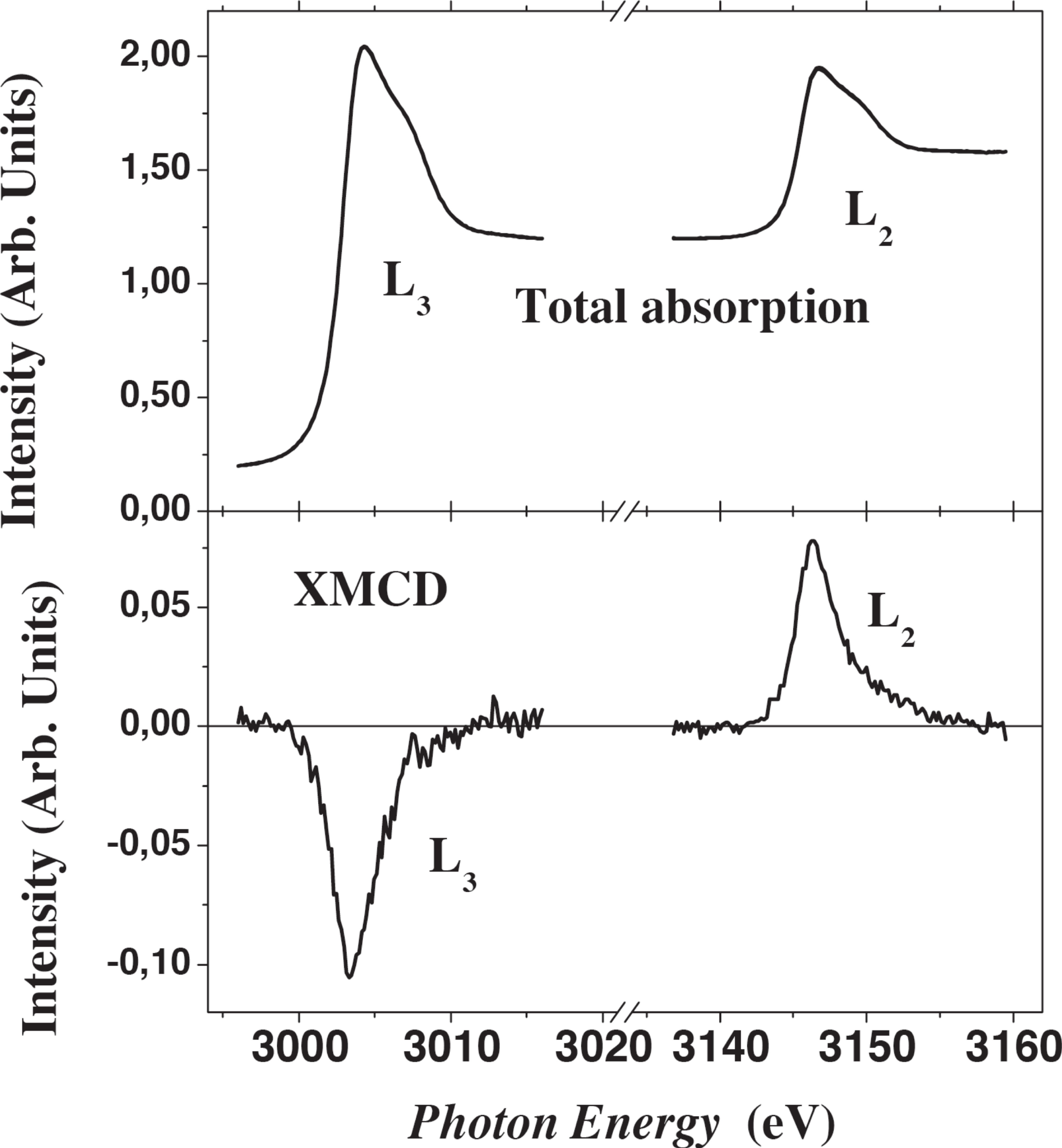}
\caption{\label{fig:xmcd}
Total absorption and XMCD spectra of 2~nm Co$_{0.5}$Rh$_{0.5}$ NPs
at the Rh L$_{3}$ and L$_{2}$ edges measured at 10K.
        }
\end{figure}

The role of $4d$ magnetic contributions has been determined experimentally  
by means of X-ray magnetic circular dichroism (XMCD) 
measurements at the L$_{2,3}$ Rh thresholds. The total absorption spectra
were recorded with the X-rays at normal incidence and 
parallel to the direction of the magnetization by detecting 
the fluorescence yield at a temperature of 10 K and a magnetic 
field of $7$ T. 
The dichroism signal has been corrected taking into account the 
incomplete polarization of the X-rays, and its accuracy has been 
checked by reversing the magnetization direction with the applied 
magnetic field. Results for 2~nm Co$_{0.5}$Rh$_{0.5}$ NPs are 
shown in Fig.~\ref{fig:xmcd}. 
The data demonstrate that the Rh atoms in Co$_{0.5}$Rh$_{0.5}$
clusters carry significant magnetic moments. 
Using the usual sum rules~\cite{thole-prl}
we derive $\mu_L /\mu_S = 0.066$ for the orbital to spin ratio. 
This is in qualitative agreement with our theoretical results 
$\mu_L /\mu_S = 0.076$ for 2nm NPs having a Rh core and an outer Co shell.
In contrast, the calculations assuming a Co core and a Rh outer 
shell yield values that are systematically far too 
small ($\mu_L/\mu_S\lesssim 0.01$). 

\begin{figure}
\includegraphics[scale=0.3]{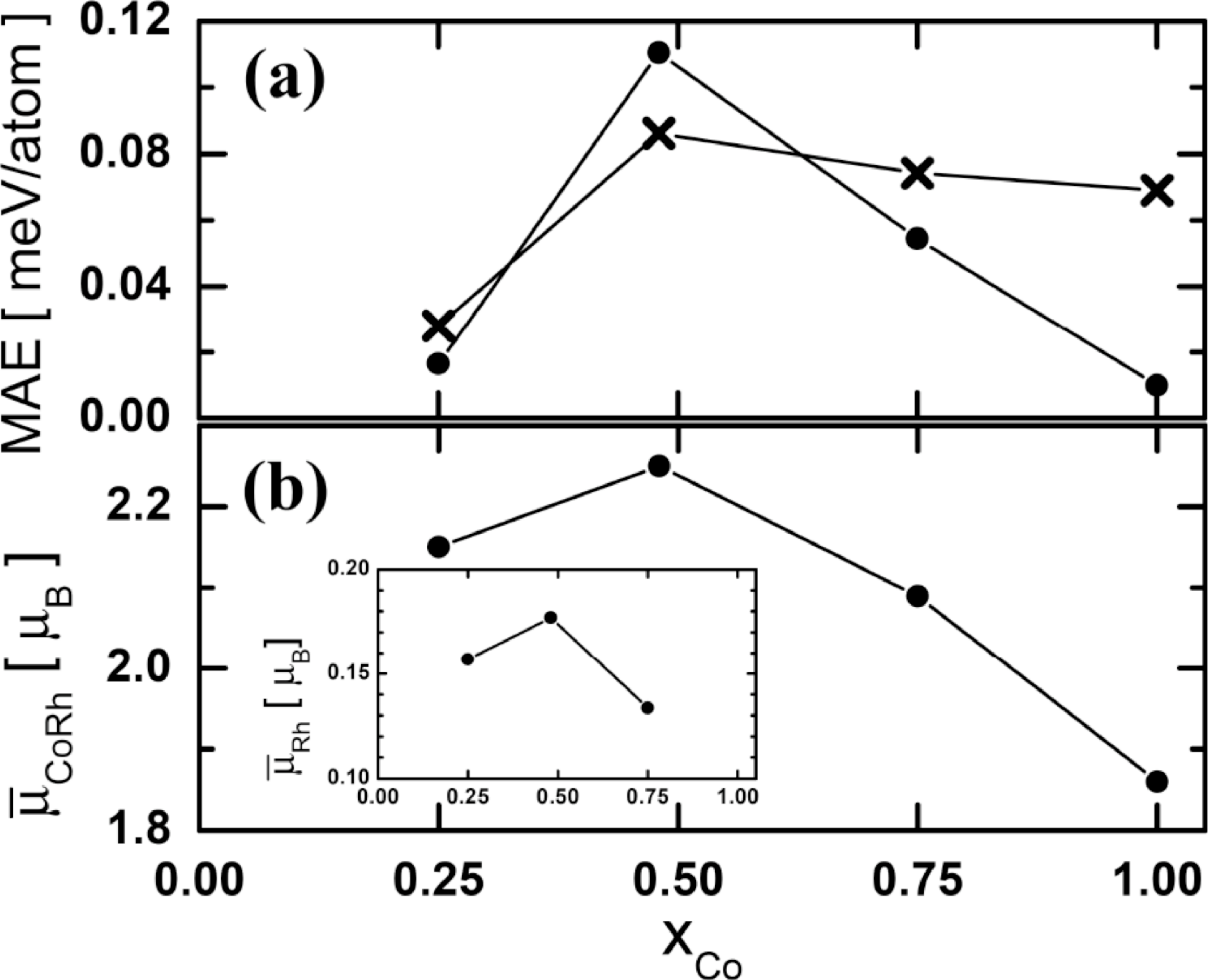}
\caption{\label{fig:mae}
(a) Magnetic anisotropy energy and (b) average magnetic moment per 
CoRh unit $\overline\mu_{\rm CoRh}$ of 2~nm 
$({\rm Co}_x{\rm Rh}_{1-x})_{N}$ as a function of Co 
concentration $x$. Crosses refer to experiment and dots to 
calculations for fcc octahedral clusters ($N = 489$ atoms). 
The inset shows the corresponding average magnetic moment 
$\overline\mu_{\rm Rh}$ induced at the Rh atoms. 
        }
\end{figure}

Finally, in Fig.~\ref{fig:mae}(a) the concentration 
dependence of the MAE of 2~nm CoRh NPs is shown. The experimental results 
were obtained 
by fitting the zero-field-cooled and field-cooled magnetization curves 
using a standard uniaxial Stoner-Wohlfarth model and
a log-normal distribution of sizes. The theoretical 
results correspond to fcc-like octahedral clusters having $N = 489$ atoms,
and a Rh core with a Co outer shell. 
A remarkable non-monotonous concentration dependence is observed.
Starting from pure Co NPs ($x=1$) and increasing the Rh content, the MAE
first increases reaching a maximum around $x = 0.5$ and then decreases
rapidly when the Co concentration is further reduced. Experiment and theory 
deliver quite consistent results, although the calculations somewhat
underestimate the measurements for $x\ge 0.75$. The microscopic origin of this 
important effect can be clarified by analyzing the concentration dependence 
of the local moments and in particular those induced at the Rh atoms.
As shown in Fig.~\ref{fig:mae}(b) the magnetic moment per Co atom and the
average magnetic moment $\overline\mu_{\rm Rh}$ at Rh atoms (inset figure) 
increase with increasing Rh content (i.e., decreasing $x$ in Co$_x$Rh$_{1-x}$) 
until the Co concentration becomes so low that the overall cluster 
magnetization breaks down. The higher magnetic susceptibility of the 
Rh clusters, as compared with Rh bulk, also explains why the optimal 
Rh concentration is larger in CoRh NPs 
($x_{max} \simeq 0.5$ for 2~nm NPs) than in macroscopic CoRh alloys 
($x_{max} \simeq 0.75$ in the bulk). The correlation between induced 
Rh moments and MAE is found to be a quite general trend. For instance, 
if fully segregated CoRh NPs are assumed, the same non-monotonous 
dependence of the induced Rh moments and MAE is obtained but with both 
values approximately half as large. These results not only demonstrate 
the optimization of the MAE and the dominant role played by $4d$ 
magnetism in $3d$-$4d$ nanoalloys. They also provide a new physical insight
on the microscopic mechanisms controlling the subtle MAE of 
these systems, that should be very useful as a guide to 
microscopic material design \cite{foot:reor}.

In conclusion, the MAE of CoRh alloys nanoparticles shows a  
non-monotonous dependence as a function of composition that opens
new possibilities of tailoring their magnetic behavior for specific
applications. The optimum $4d$/$3d$ content is found to be size 
dependent, actually higher in NPs as compared to bulk alloys. 
Combining magnetometry and XMCD experiments and theoretical
calculations we achieve a consistent
physical picture of this important effect, which is dominated by 
the contributions of the induced Rh magnetic moments and by their 
non-trivial dependence on the local and chemical environment of the atoms. 
The observed trends and the microscopic understanding derived 
in this work ---for instance, concerning the correlations 
between the induced Rh moments, MAE, and chemical order--- 
are expected to be generally applicable to the broad family of 
$3d$-$4d$ and $3d$-$5d$ nanoalloys. They should therefore provide 
a very useful guide towards a knowledge-based quantum design of 
magnetic nanostructures. Among the exciting immediate perspectives one could
mention, for example, the possibility of manipulating the kinetics of 
the synthesis process to produce different core-shell arrangements 
or even multishell nanoalloys in order to exploit the remarkable 
$3d$-$4d$ and $3d$-$5d$ interface effects.

Helpful discussions with Profs.~M.-J.~Casanove and B.~Chaudret 
are gratefully acknowledged. This work has been supported by CONACyT 
(Grant No.\ 62292 and U5065), CNRS, INSAT and DAAD. 


\begin{thebibliography}{99}
\bibitem{theo2b}
Further details will be published elsewhere.
\bibitem{foot:reor}
Extended calculations including the effects of alloy disorder 
yield qualitatively similar results, in particular concerning the 
non-monotonous concentration dependence of the MAE. They confirm 
the dominate role of Rh moments. In addition, remarkable magnetic reorientation 
transitions as a function of composition are observed 
\protect\cite{theo2b}.
\end{thebibliography}
\end{document}